\documentclass[12pt,article,nofootinbib]{revtex4}
\usepackage{pdfpages}
\usepackage{epsf}
\usepackage{subfigure}
\usepackage{amsmath}
\usepackage{eqnarray}
\usepackage{epstopdf}
\usepackage{mathrsfs}
\usepackage{natbib}
\usepackage{amssymb}
\usepackage{dcolumn}
\usepackage{graphicx}
\usepackage{color} 
\usepackage{hyperref}
\usepackage{enumerate}
\usepackage{float}

\usepackage{multirow}
\usepackage{blindtext}

\makeatletter
\newcommand*{\rom}[1]{\expandafter\@slowromancap\romannumeral #1@}
\makeatother
\def\be{\begin{equation}}
\def\ee{\end{equation}}
\def\ba{\begin{eqnarray}}
\def\ea{\end{eqnarray}}

\graphicspath{{images/}}
\begin{document}

	\title{Hawking Temperature for 4D-Einstein-Gauss-Bonnet Black Holes from uncertainty principle}
	\author{Sara Azizi}
	\email{sarahazizi@email.kntu.ac.ir}
	\affiliation{Department of Physics, K. N. Toosi University of Technology, P.O. Box 15875-4416, Tehran, Iran}
	\author{Sareh Eslamzadeh}
	\email{S.Eslamzadeh@stu.umz.ac.ir}
	\affiliation{Department of Theoretical Physics, Faculty of Basic Sciences, University of Mazandaran, Babolsar, IRAN }
	\author{Javad T. Firouzjaee}
	\email{firouzjaee@kntu.ac.ir}
	\affiliation{Department of Physics, K. N. Toosi University of Technology, P.O. Box 15875-4416, Tehran, Iran}
	\affiliation{School of Physics, Institute for Research in Fundamental Sciences (IPM), P.O. Box 19395-5531, Tehran, Iran}
	\author{Kourosh Nozari}
	\email{knozari@umz.ac.ir}
	\affiliation{Department of Theoretical Physics, Faculty of Basic Sciences,
		University of Mazandaran, P. O. Box 47416-95447, Babolsar, IRAN }
	\begin{abstract}
		 		
		Inspired by string theory, Heisenberg's uncertainty principle can be generalized to include the photon-electron gravitational interaction, which leads to the Generalized Uncertainty Principle (GUP). Although GUP considers gravitational uncertainty at the minimum fundamental length scale in physics, it does not consider the effects of spacetime curvature on quantum mechanical uncertainty relations. The Extended Uncertainty Principle (EUP) is a generalization of Heisenberg's Uncertainty Principle that, unlike the GUP, applies to large length scales. GEUP is also a linear combination of EUP and GUP that creates minimal uncertainty on large length scales. The Einstein-Gauss-Bonnet theory (EGB) can be considered as one of the most promising candidates for modified gravity.  In this paper, by using GUP, EUP, and GEUP, we intend to obtain the Hawking temperature of a four-dimensional EGB black hole in the asymptotically flat and (Anti)-de Sitter spacetime. We show that coupling constant, cosmological constant, mass, and radius significantly affect Hawking temperature and decrease or increase Hawking temperature depending on the chosen horizons.\\
		
{\bf Keywords:} Generalized Uncertainty Principle, Extended Uncertainty Principle, Einstein-Gauss-Bonnet theory, (Anti)-de Sitter spacetime, Hawking temperature, Cosmological constant.
	\end{abstract}

	\maketitle
	\newpage
	\tableofcontents
			
	\section{Introduction}
	
	General relativity (GR) and quantum field theory (QFT) are leading theories that describe gravitational interactions and phenomena at the atomic and subatomic scales, respectively. Both of these theories are in good agreement with the experiments. In GR, Einstein predicted the existence of black holes, which was eventually proven by observations made on the gravitational wave channel by LIGO / VIRGO \cite{1,2,3}. In addition, the existence of black holes was once again confirmed by the Event Horizon Telescope, which recorded the first image of a black hole shadow \cite{4,5}. On the other hand, QFT also accurately predicts constants of many microscopic structures using experimental measurements \cite{6,7,8}. Although GR and QFT are in good agreement with experiments, the current perspective of gravity on the Planck scale does not provide a unified description of theoretical physics. Using Dyson's power-counting method, it can be shown that GR is not a theory that can be re-normalized \cite{9}. Therefore, alternative theories of gravity are introduced that can be re-normalized. Correction of quadratic-order curvature in GR may give rise to renormalizable gravity theories \cite{10}.
	It is expected that the problem of singularity formation in GR will be solved with the help of such theories of gravity \cite{11}. 
	In GR, the Einstein-Hilbert action is modified to include the so-called Gauss-Bonnet term, which the resulting theory is called the Einstein-Gauss-Bonnet gravity (EGB) \cite{12}. The EGB theory is an alternative theory of gravity that could be one of the most promising candidates for alternative theories of gravity. The action of EGB gravity in a D-dimensional spacetime with a positive cosmological constant is as follows 
	
	\begin{equation}\label{eq:area}	
		S_{EGB}=S_{EH} + S_{GB} = \frac{1}{16\pi G_{D}} \int d^{D}x \sqrt{-g} [R-2\Lambda + \alpha \mathcal{L}_{GB}]	, 
	\end{equation}
	
	\noindent where $ G_{D} $ is the D-dimensional Newton’s constant, $ R $ is the Ricci scalar, $ \Lambda $ is the cosmological constant, and $ \alpha $ is the Gauss-Bonnet coupling constant. The Gauss-Bonnet term has been introduced as follows
	
	\begin{equation*}\label{eq:area}	
		\mathcal{L}_{GB} = R_{\mu\nu \rho\sigma}R^{\mu\nu\rho\sigma} - 4R_{\mu \nu}R^{\mu\nu} + R^{2}.
	\end{equation*}
	
	In the past, it was believed that EGB gravity was nontrivial only in the 4 + 1 dimensions and higher dimensions, and would therefore only be used in high-dimensional models. In fact, EGB was reduced to a topological level term in the lower dimensions \cite{14}. Although this expression was quadratic in the Riemann tensor and the Ricci tensor, terms with more than two partial derivatives of the metric would be omitted, and the Euler-Lagrange equations in the metric would be converted to second-order quasi-linear partial differential equations. As a result, like $f(R)$ gravity, there would be no other dynamical degrees of freedom \cite{15}.
	
	In a recent paper, \cite{16}, D. Glavan and C. Lin argued that if the Gauss-Bonnet coupling constant scale in EGB theory can be changed to $D>4$, then if the computations were made when $D\rightarrow4$, the contribution of the Gauss-Bonnet expression in the 4-dimensions can be obtained nontrivial. The general metric equation and symmetric spherical solution of the EGB black hole in 4-dimensional spacetime are as follows
	
	\begin{equation}\label{eq:area}	
		ds^{2}= - f(r) dt^{2} + \frac{dr^{2}}{f(r)}+r^{2}	d\Omega^{2}_{2},
	\end{equation}
	
	\begin{equation}\label{eq:area}	
		f(r) = 1 + \frac{r^{2}}{2\alpha} \left(1- \sqrt{1 + \dfrac{8 \alpha M }{r^{3}} + \dfrac{4 \alpha \Lambda}{3}}\right), 
	\end{equation}
	
	\noindent where $ d\Omega^{2}_{2} $ is the line element of the two-dimensional unit sphere $ S^2 $. The 4D-EGB gravity has been studied in various areas such as obtaining observational constraints on regular 4-dimensional EGB gravity \cite{17}, investigating the power spectra of the Hawking radiation, and greybody factor of a massless scalar \cite{18}, etc. \cite{19,20, 21,22, 23, 24, 25, 26, 27, 28}. In Refs. \cite{22, 23}, the authors obtained the energy emission rate and greybody factor in various fields, including Dirac, electromagnetic and gravitational field, using the asymptotically flat 4D-EGB theory, and calculated the lifetime of the black hole for different values of coupling constants. In addition, in the Refs. \cite{25,26, 27, 28}, the evaporation process of the AdS EGB black holes in cases with D dimensions and different properties of the charged black holes in AdS 4D-EGB spacetime at various coupling constants have also been investigated.
	
	In addition to the success of this theory, it is worth mentioning some of its criticisms. Some authors believe that the 4D-EGB solution developed by Glavan and Lin needs to be modified \cite{29,30, 31, 32, 33, 34}. In Ref. \cite{29}, the authors obtained a set of field equations that can be written in closed form in four dimensions by inserting a counter term in the action. The method presented in this reference leads to a theory of gravity that includes an additional scalar gravitational degree of freedom. In the Ref. \cite{30}, Kobayashi uses the Kaluza-Klein reduction to promote the warp factor of the internal space to a scalar degree of freedom and shows that it is possible to describe the regularized Lovelock gravity through a certain subclass of Horndeski theory.
	In addition, Aoki, Gorji, and Mukayama recently presented a consistent theory of 4D-EGB gravity with two degrees of freedom based on reasonable assumptions. According to Hamiltonian formalism, they showed that it is not possible to obtain a consistent gravitational theory with two degrees of freedom when we incline the $ (d+1) $-dimensional EGB gravity to $ d \rightarrow 3 $. They pointed out that either an extra degree of freedom has to be introduced into the theory or temporal diffeomorphism has to be violated to obtain a consistent $ (3+1) $-dimensional theory \cite{31}.
	
	One of the main features of a black hole is its temperature which is characterized through the Hawking process. The origin of this radiation is the creation of quantum particles in the background vacuum due to the Uncertainty Principle \cite{firouzjaee2}. The Heisenberg Uncertainty Principle (HUP) forms the cornerstone of quantum physics. Heisenberg showed that a microscopic particle's exact position and momentum, the size of an electron, could never be measured precisely  simultaneously \cite{36}. In fact, this principle constitutes a fundamental limitation for the measurement of classical trajectories at the atomic or sub-atomic scales. However, in HUP, there is no absolute maximum or minimum uncertainty for momentum and position. HUP does not consider the effects of quantum gravity on photon interactions and spacetime curvature \cite{37}. Calculations performed in recent research in the fields of non-commutative geometries \cite{38}, string theory \cite{39}, black hole physics \cite{40}, Hawking radiation spectrum \cite{Firouzjaee:2015wps}, and quantum gravity \cite{42} show the existence of a minimum length \cite{43}. As a result of these arguments, a generalization of the Heisenberg uncertainty principle was introduced, called the generalized uncertainty principle (GUP), which predicts the minimum length value on the Planck length scale \cite{44,45}. The absolute minimum in the position uncertainty restricts the black hole evaporation and, unlike the HUP, prevents Hawking temperature divergence. Increasing Hawking temperature by GUP with absolute minimum length ($ r_{+}=2al_{p} $) leads to a faster decay of the Schwarzschild black hole in any dimension \cite{46, 47}. 
	
	Similar to HUP, GUP does not have a precise limit for maximum position uncertainty, so it cannot be used on large length scales such as de Sitter or Anti-de Sitter space. In HUP or GUP, the Hawking temperature of Schwarzschild black holes cannot be reproduced in the (Anti)-de Sitter space \cite{37}.
	The extended uncertainty principle (EUP) covers large scales and predicts the minimum amount of measurable momentum \cite{48}. In the (Anti)- de Sitter background, the EUP contains a correction that corresponds to the (Anti)- de Sitter radius. This expression can also be derived from (Anti)- de Sitter spacetime geometry. Using EUP, one can accurately obtain Hawking temperatures in (Anti)- de Sitter black holes, unlike HUP and GUP \cite{49}.
	In Ref. \cite{37}, the EUP is generalized to include position uncertainties and calculate Hawking temperatures correctly. The generalized extended uncertainty principle (GEUP) can be obtained from the linear combination of GUP and EUP to simultaneously achieve high energy and large-scale changes \cite{49}.
	
	In this paper, we obtain the Hawking temperature of the 4D-EGB Schwarzschild black hole using the GUP, EUP, and GEUP and investigate how the coupling constant affects the Hawking temperature. In addition, we study Hawking temperature behavior based on the mass, cosmological constant, and radii. 
	
	This article is organized as follows: In Section II, we obtain the Hawking temperature through the GUP for 4D-EGB black holes in asymptotically flat space. Sections III and IV  are devoted to obtaining the Hawking temperature in Anti-de Sitter and de Sitter space for the 4D-EGB black hole using EUP and GEUP, respectively, and investigating the mass, the cosmological constant, and coupling constant effects on Hawking temperature. Finally, Section V summarizes the works done in this article.\\

	\section{The GUP and Hawking temperature in an asymptotically flat 4D-EGB black hole}
	In this section, we recall the generalized uncertainty principle and then obtain the Hawking temperature for the asymptotic flat 4D-EGB black hole. We consider the following form of GUP
	
	\begin{equation}\label{eq:area}	
		\Delta x_{i} \Delta p_{j} \geq \hbar \delta _{ij} \left[ 1+ a ^{2} l^{2}_{P} \frac{(\Delta p_{j})^{2}}{\hbar ^{2}} \right],
	\end{equation}
	
	\noindent where $ x_{i} $ and $ p_{j} $  are spatial and momentum coordinates, respectively, $ l_{p} $ is Planck length, and $a$ is a dimensionless real constant of order one. GUP has a variety of representations, and Eq. (4) is just one of its representations, so the precise value of $ a $ depends on the chosen model \cite{50}. The second term on the right-hand side creates an absolute minimum on position uncertainty
	
	\begin{equation}\label{eq:area}	
		\Delta x_{i} \geq 2 a l_{P}, 
	\end{equation}
	
	\noindent which non-trivially gives a minimal measurable length.
	The uncertainty in the momentum is obtained from Eq. (4) as
	
	\begin{equation}\label{eq:area}	
		\frac{\hbar \Delta x_{i}}{2 a^{2} l^{2}_{P}} \left[ 1 - \sqrt{1-\frac{4a^{2}l^{2}_{P}}{(\Delta x_{i})^{2}}} \right] \leq \Delta p_{i} \leq \frac{\hbar \Delta x_{i}}{2 a^{2} l^{2}_{P}} \left[ 1 + \sqrt{1-\frac{4a^{2}l^{2}_{P}}{(\Delta x_{i})^{2}}} \right].
	\end{equation}
	
	The left-hand inequality produces small corrections to the Heisenberg’s uncertainty principle for $ \Delta x_{i} \gg a l_{P} $. For the corrections in the GUP formula to be viable, the inequality on the right-hand side cannot be arbitrarily large.
	We now derive the Hawking temperature from the generalized uncertainty principle and examine the general properties of black holes in flat spacetime. In flat spacetime, the symmetric spherical solution of the 4D-EGB black hole is as follows
	
	\begin{equation}\label{eq:area}	
		f(r)= 1+ \frac{r^{2}}{2\alpha} \left[ 1 \pm \sqrt{1+ \frac{8\alpha M}{r^{3}}}\right],  
	\end{equation}
	
	\noindent where the cosmological constant in flat space vanishes in Eq. (3), and $ M $ is a positive gravitational mass. From the metric $ f(r) $, two horizons $ r_{\pm} $ can be obtained
	
	\begin{equation}\label{eq:area}	
		r_{\pm} = M \left[ 1 \pm \sqrt{1-\frac{\alpha}{M^{2}}} \right], 
	\end{equation}
	
	\noindent where $r_{+}$ is the black hole horizon, and $r_{-}$ is the white hole horizon. According to Eq. (8), there is a critical mass that depends on the coupling constant $\alpha$. When $M>M_{cri}=\sqrt{\alpha}$, the black hole has two horizons: the white hole horizon and the black hole horizon. When $M=M_{cri}$, the black hole horizon, and the white hole horizon will merge, and if $M<M_{cri}$, there will be no horizon \cite{13}.
	Then, based on the Ref. \cite{37}, we model a 4-dimensional black hole like a black box with radius $ r $, the uncertainty in the position of a particle propagated by the radius of the  horizon $ r_{\pm} $ is equal to 
	
	\begin{equation}\label{eq:area}	
		\Delta x_{i} \approx r_{\pm},
	\end{equation}
	\noindent where
	
	\begin{equation}\label{eq:area}	
		r_{\pm} = M \pm \sqrt{M^{2} - \alpha}.
	\end{equation} 
	
	We ignore the mass of the emitted particle, so the energy uncertainty of the particle is equal to $ E $, which can be considered as the characteristic temperature of Hawking radiation. If we assume $ E $ is equal to left inequality in relation (6), the Hawking temperature can be obtained as follows
	
	\begin{equation}\label{eq:area}	
		T_{GUP}= \left( \frac{d-3}{4\pi} \right) \frac{\hbar r_{+}}{2a^{2} l^{2}_{P}} \left[ 1-\sqrt{1-\frac{4a^{2} l^{2}_{P}}{r^{2}_{+}}} \right],
	\end{equation}
	
	\noindent where $ ``(d - 3) / 4\pi" $ is the calibration factor and we set $\hbar = l_{P}= 1$ and $ d = 4 $ here \cite{51} to find
	
	\begin{equation}\label{eq:area}	
		T_{GUP}=\frac{r_{+}}{2a^{2}4\pi} \left[ 1-\sqrt{1-\frac{4a^{2}}{r^{2}_{+}}} \right],
	\end{equation}
	
\begin{figure}[H]
		\centering
		\includegraphics[height=7 cm,width=7.5 cm]  {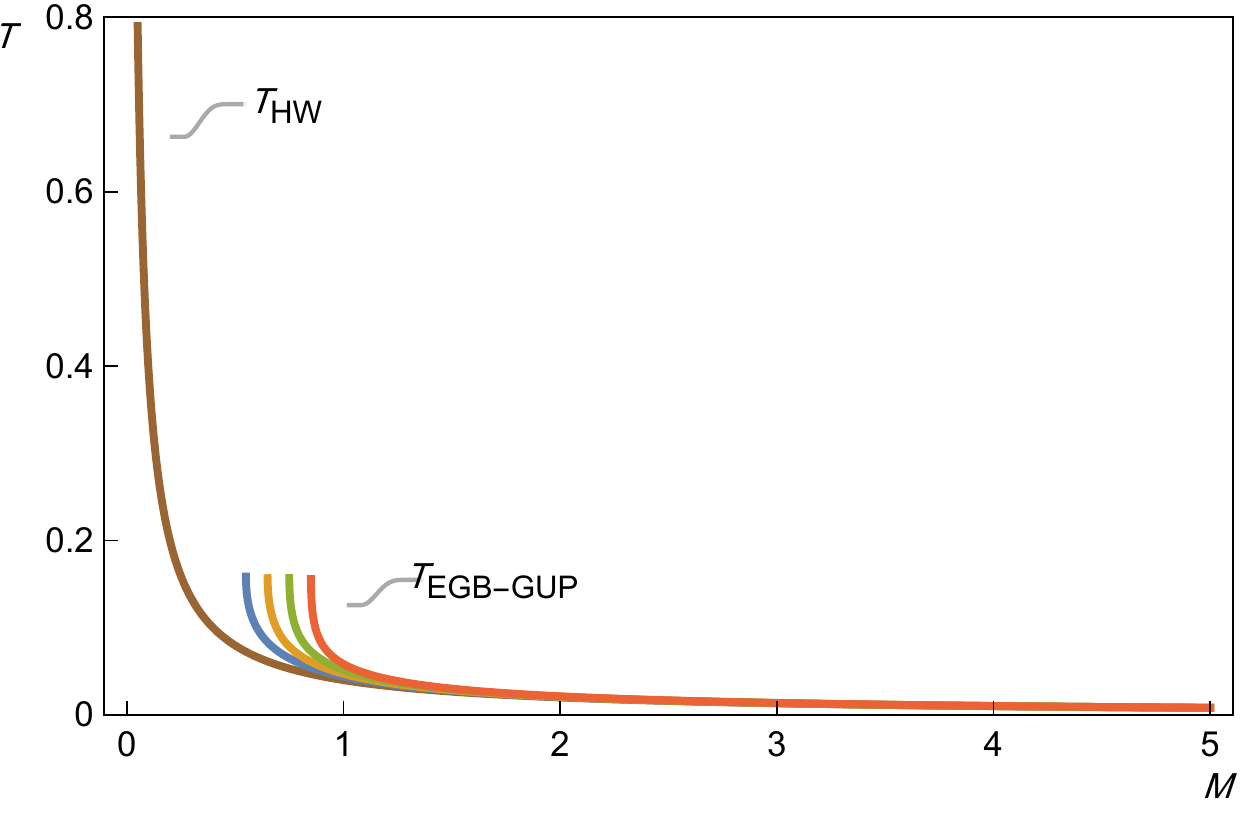}
		\hspace*{0.1cm}
		\includegraphics[height=7 cm,width=7.5 cm] {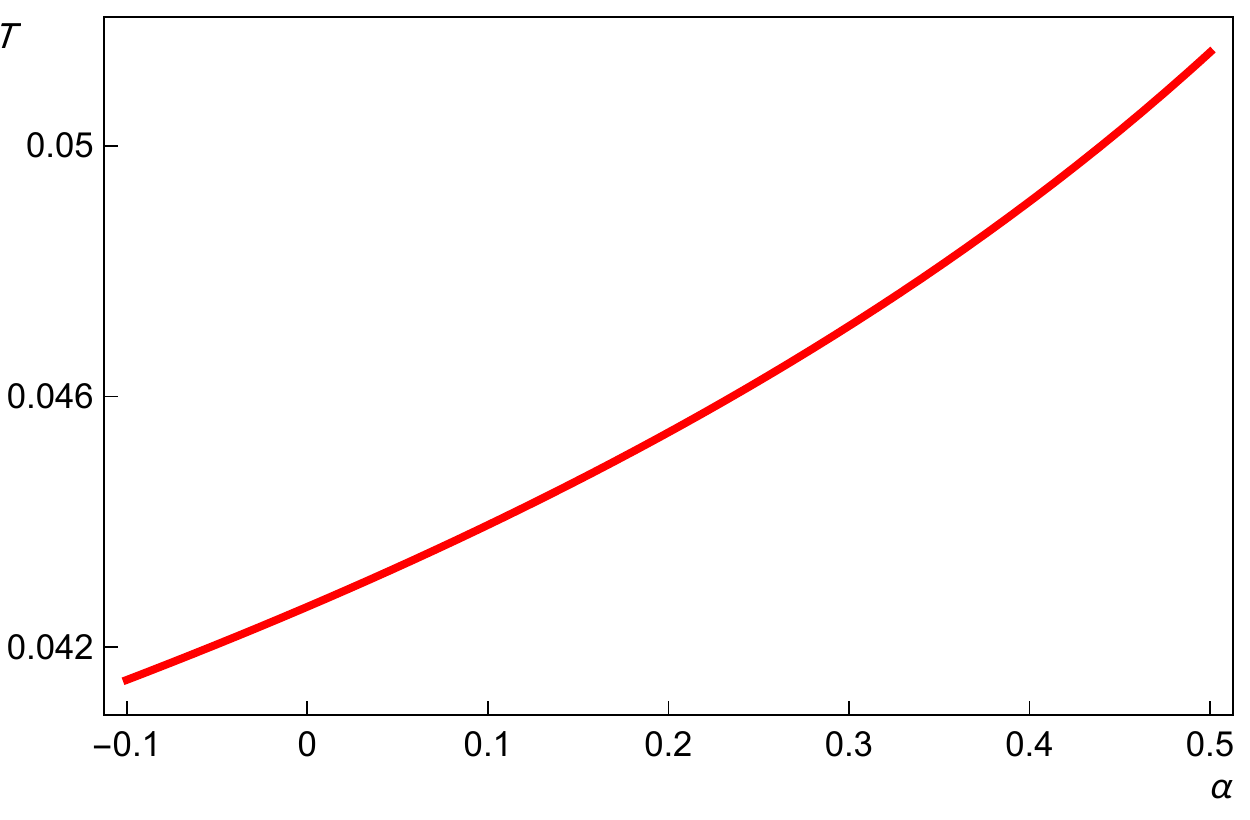}
		\hspace*{0.1cm}
		\includegraphics[height=7 cm,width=7.5 cm] {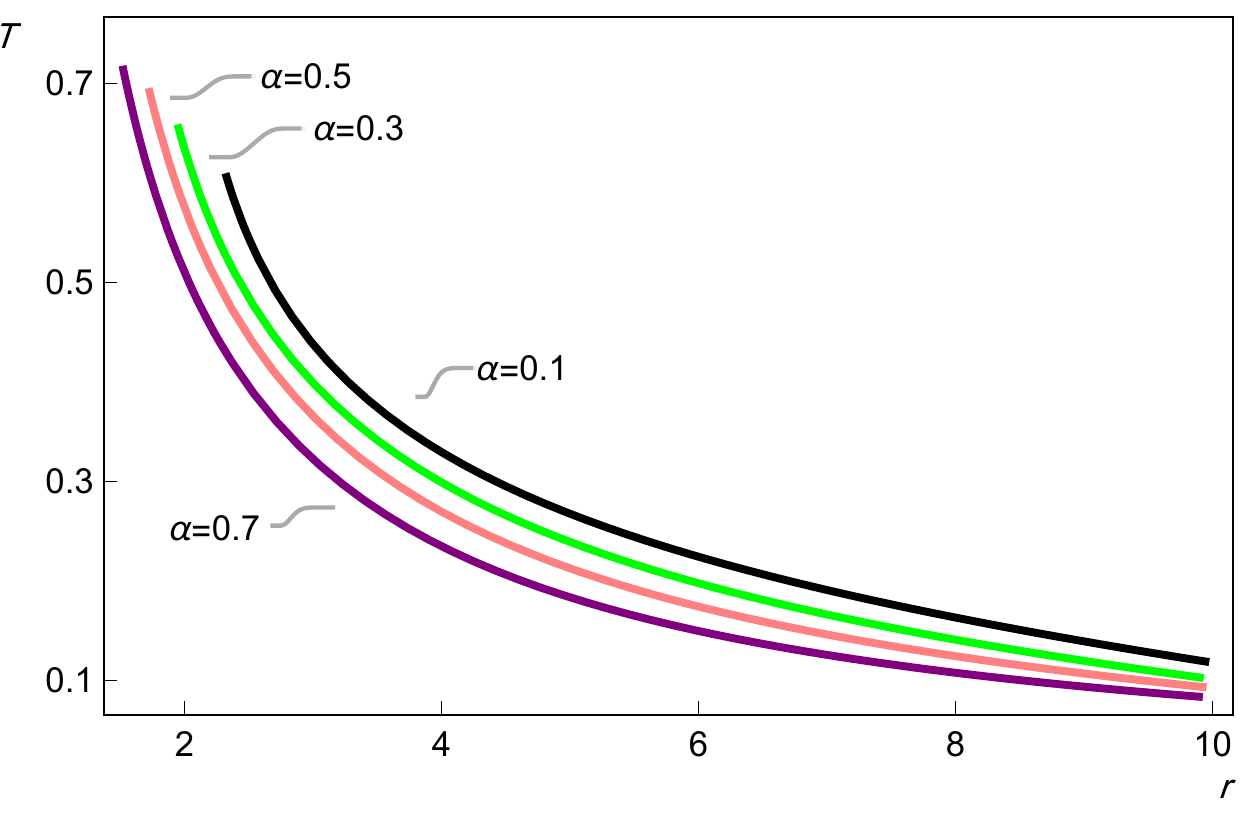}
		\caption{\scriptsize{Hawking temperature of black hole horizon versus mass, coupling constant, and radius for asymptotic flat 4D-EGB black hole. 
		Left diagram: Hawking temperature has been plotted for different masses from left to right with $ \alpha=0.1, 0.3, 0.5, 0.7$. Right diagram: Hawking temperature versus coupling constant has been plotted with $ M = 1 $. Bottom diagram: Hawking temperature versus radius has been plotted with $ \alpha = 0.1, 0.3, 0.5, 0.7 $.}}		
		\label{figure_1}
	\end{figure}

	We treat the Hawking temperature behavior versus the mass and coupling constant changes by replacing the black hole horizon radius (10) in Eq. (12) and plotting the $ T - M $ and $ T - \alpha $ diagrams. 
	The first plot on the left panel of Fig. \eqref{figure_1} shows the Hawking temperature and $ T_{EGB-GUP} $ with different coupling constants. Hawking temperature increases and diverges without GUP as the mass decreases. It is illustrated that $ T_{EGB-GUP} $ increases with decreasing mass, and $ T_{EGB-GUP} $ is limited at rest-mass remnant and black hole evaporation ends \cite{52, 53, 54, 55}.\\

	
	 In Ref. \cite{13}, the $ T_{EGB}-M $ is plotted and this plot shows that EGB corrects the temperature behavior of the black hole at a small radius and stops temperature divergence. Although the EGB predicts zero temperature for the end of the black hole evaporation and has a mass remnant, its temperature is zero. However, comparing this plot with the $ T-M $ in Fig. \eqref{figure_1}, it can be concluded that when the GUP enters the EGB, it stops the evaporation of the black hole where the temperature has not yet reached zero. In fact, a black hole with a finite mass within the Planck mass range achieves a limited temperature \cite{56}.
	
	As can be seen from the $ T-\alpha $ plot, the Hawking temperature increases with increasing $ \alpha $. In EGB-GUP, if we consider a more significant $ \alpha $, we will obtain a larger temperature for the remnant mass anticipated at the end of the black hole evaporation.
	
	When we plot the $ T-r $ with different $ \alpha $, one can see that the Hawking temperature change with the radius of the black hole is as expected, and when $ r $ decreases, the Hawking temperature increases. Having a minimum length due to GUP will limit the Hawking temperature and prevents it from diverging at small lengths \cite{37}. The effect of the coupling constant on temperature is negligible; however, increasing the positive coupling constants causes a slight decrease in Hawking temperature. \\

	\section{Hawking temperature through EUP and GEUP in an asymptotically Anti-de Sitter 4D-EGB black hole}
	As mentioned earlier, GUP cannot be used in large-length spaces such as (Anti)-de Sitter spaces. Therefore, we obtain the Hawking temperature from the uncertainty relation in such spaces by extending Heisenberg's uncertainty principle to large-length scales \cite{37, 57}. 
	The extended uncertainty relation is
	
	\begin{equation}\label{eq:area}	
		\Delta x_{i} \Delta p_{j} \geq \hbar \delta_{ij} \left[ 1 + \beta^{2} \frac{(\Delta x_{i})^{2}}{l^{2}} \right], 
	\end{equation}
	
	\noindent where $ l $ is the characteristic large-length scale and $ \beta $ is a dimensionless real constant of order one \cite{49}. The EUP in AdS spacetime requires an absolute minimum of momentum uncertainty
	 
	\begin{equation}\label{eq:area}
		\Delta p_{i} \geq \frac{\hbar}{\Delta x_{i}} + \frac{\hbar \beta^{2} \Delta x_{i}}{l} \geq \frac{2 \hbar \beta}{l}.
	\end{equation}
	
	\noindent Furthermore, GEUP is a combination of GUP and EUP that includes both a minimum length scale and a minimum momentum scale
	\begin{equation}\label{eq:area}	
		\Delta x_{i} \Delta p_{j} \geq \hbar \delta_{ij} \left[ 1 + a^{2} l^{2}_{P} \frac{(\Delta p_{j})^{2}}{\hbar^{2}} + \beta^{2} \frac{(\Delta x_{i})^{2}}{l^{2}} \right]. 
	\end{equation}
	
	\noindent In GEUP, there is also an absolute minimum for the position uncertainty due to the Planck length
	
	\begin{equation}\label{eq:area}	
		(\Delta x_{i})^{2} \geq \frac{4a^{2}l^{2}_{P}}{1-4a^{2}l^{2}_{P} \beta^{2} / l^{2}},
	\end{equation}

	\noindent and the EUP here also creates an absolute minimum for the momentum uncertainty
	
	\begin{equation}\label{eq:area}	
		(\Delta p_{i})^{2} \geq \frac{4 \hbar^{2} \beta^{2} / l^{2}}{1-4a^{2}l^{2}_{P} \beta^{2} / l^{2}}. 
	\end{equation}
	
	Studies show that the existence of minimum length always raises the temperature and accelerates evaporation of a black hole, both in flat space and in (Anti)-de Sitter space \cite{46, 47, 58}.
	In this section, we first obtain the Hawking temperature of the 4D-EGB black hole for the Anti-de Sitter spacetime. In the following, we will investigate the effect of $ \alpha $, $ \Lambda $, and $ M $ on the Hawking temperature of the Anti-de Sitter 4D-EGB black hole, and we will analyze the effect of minimum length in GEUP on the black hole, as well. 
	Considering the solution (2) in the AdS space, the metric function is as follows
	
	\begin{equation}\label{eq:area}	
		f(r) = 1 + \frac{r^{2}}{2\alpha} \left (1- \sqrt{1 + \dfrac{8 M \alpha  }{r^{3}} - \dfrac{4 \alpha \Lambda}{3}} \right ).  
	\end{equation}
	
	Here the cosmological constant $\Lambda = - (d-1)(d-2) / 2{l^{2}_{_{AdS}}} $ \cite{37}. We adopted the negative sign into the metric $ f(r) $, so the $ \Lambda $ would be non-zero and positive. In Ref. \cite{13}, $ f(r) $ versus $ r $ is illustrated. According to this diagram, as in the previous section, there is a critical mass. If $ M \textgreater M_{cri} $, the black hole will have two horizons: a black hole horizon and a cosmological horizon, respectively.
	The AdS 4D-EGB black hole horizons can be found by obtaining the roots of the metric Eq. (18)
	
	\begin{equation}\label{horizonds}
		r=\pm\frac{1}{2} \Big[\sqrt{\eta}\mp\sqrt{\frac{\eta-12 M}{\eta\Lambda}}\Big],
	\end{equation}
	
	\noindent where
	\begin{equation}
		\eta=\frac{2}{\Lambda}+\frac{3\times 2^{1/3}(1+4\alpha \Lambda)}{\Lambda \xi^{1/3}}+\frac{\xi^{1/3}}{3\times2^{1/3} \Lambda},
	\end{equation}
	\noindent and
	\begin{equation}
		\xi=972 \Lambda M^2 -648 \alpha \Lambda+54+\sqrt{(972 \Lambda M^2 -648 \alpha \Lambda+54)^2-4(9+36 \alpha \Lambda)^3}.
	\end{equation}

\begin{figure}[H]
		\centering
		\includegraphics[height=7 cm,width=7.5 cm]  {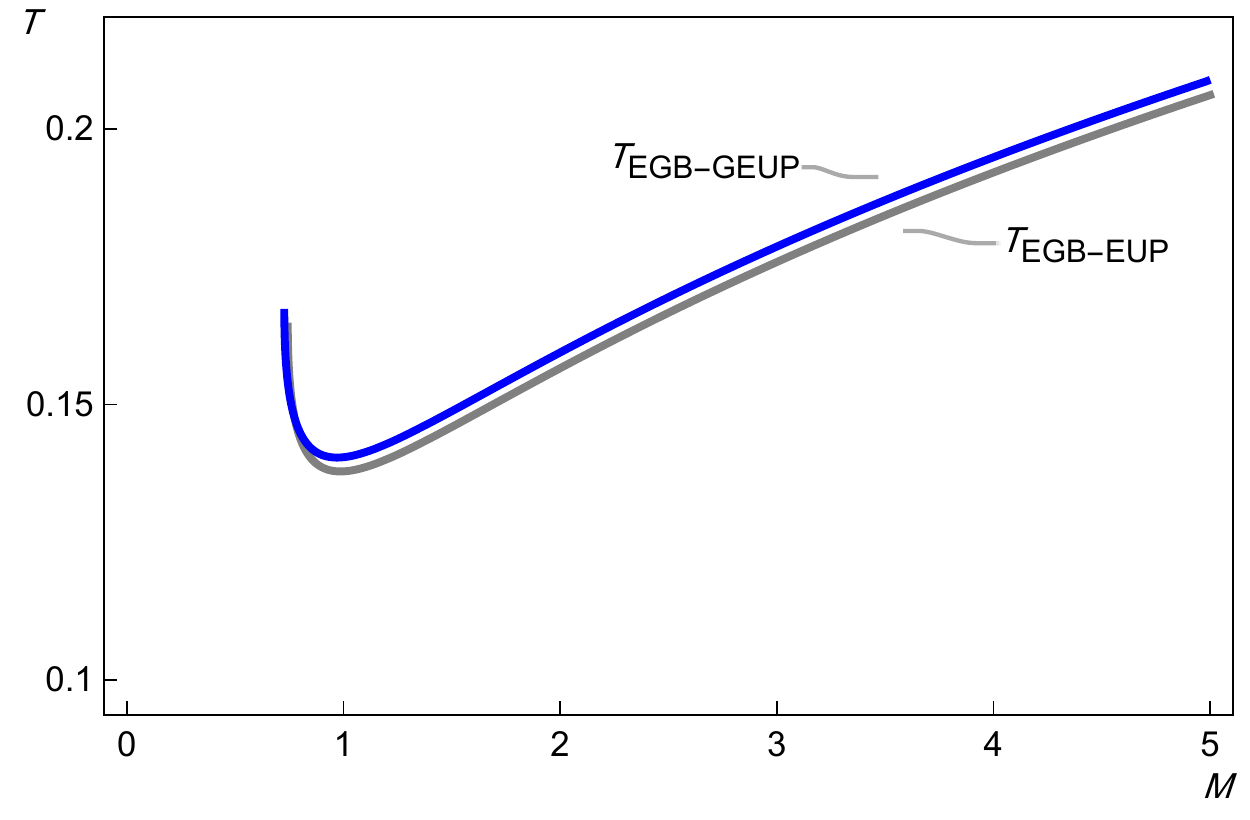}
		\hspace*{0.5cm}
		\includegraphics[height=7 cm,width=7.5 cm] {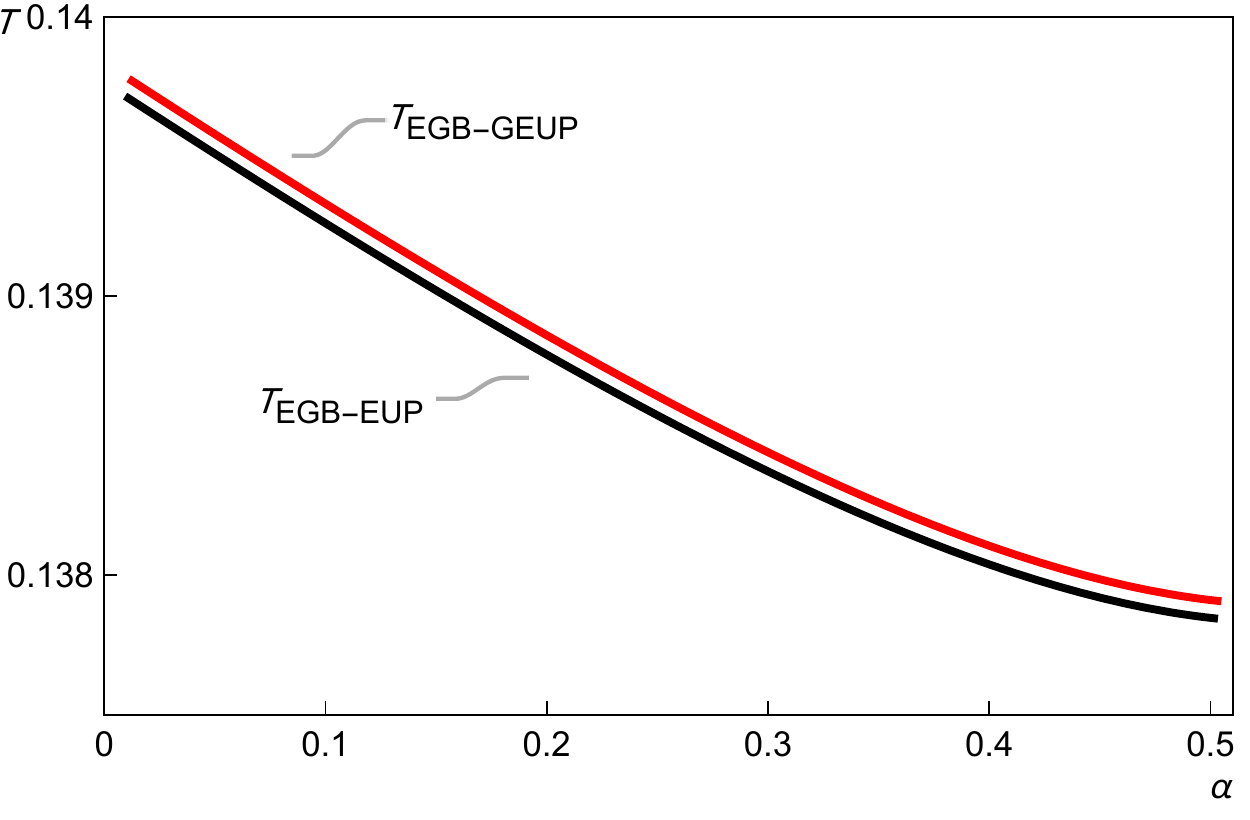}
		
		\includegraphics[height=7 cm,width=7.5 cm] {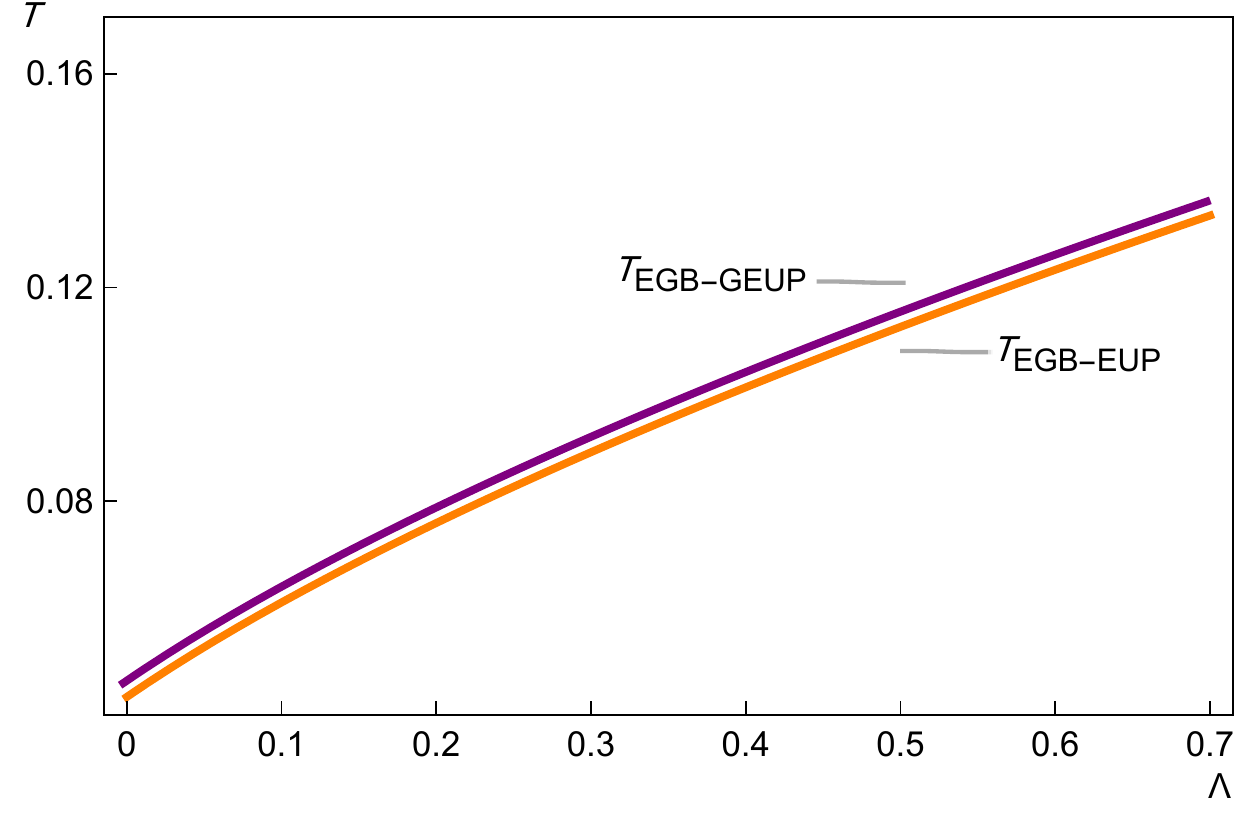}
		\hspace*{0.5cm}
		\includegraphics[height=7 cm,width=7.5 cm] {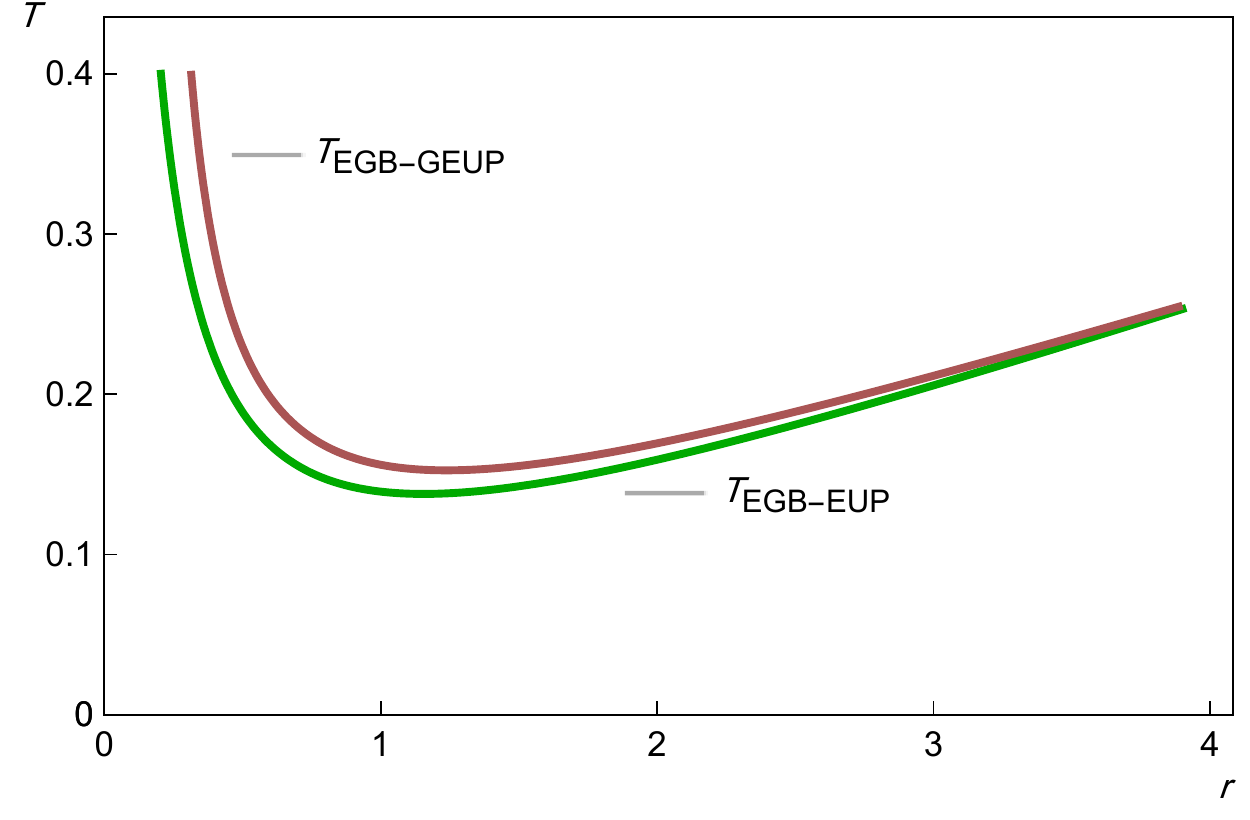}
		\caption{\scriptsize{Hawking temperature of cosmological horizon versus mass, coupling constant, cosmological constant and radius for AdS 4D-EGB black hole. From the top left to right, Hawking temperature has been plotted versus masses and coupling constant with $ \alpha=0.5$ and $ M = 1$, respectively. From bottom left to right, Hawking temperature versus cosmological constant and radius has been illustrated with $ \alpha = 0.3 $. Here in all cases $ l_{AdS}=2 $ and $ \Lambda=0.7 $.}}
		\label{figure_2}
	\end{figure}

	Of the four roots of $ r $, only two can be identified as the black hole horizon and the cosmological horizon. Using the same method as in the previous section, one can obtain the general $ T_{EGB-EUP} $ and $ T_{EGB-GEUP} $ equations with $ T \approx \Delta p_{i} $ (with $ c=1 $)
	
	\begin{equation}\label{eq:area}	
		T_{EUP(AdS)} = \left(\frac{d-3}{4\pi}\right) \hbar \left[ \frac{1}{r_{+}} + \left( \frac{d-1}{d-3} \right) \frac{r_{+}}{l^{2}_{_{AdS}}} \right], 
	\end{equation}
	
	\begin{equation}\label{eq:area}	
		T_{GEUP(AdS)} = \left( \frac{d-3}{4\pi}\right) \frac{\hbar r_{+}}{2a^{2}l^{2}_{P}} \left[ 1 - \sqrt{1 - \frac{4a^{2} l^{2}_{P}}{r^{2}_{+}} \left[1+ \left(\frac{d-1}{d-3} \right) \frac{r^{2}_{+}}{l^{2}_{AdS}}\right]}\right],  
	\end{equation}

	\noindent where $``(d - 3) / 4\pi"$ is the calibration factor, $ \beta = \sqrt{(d - 1)/(d - 3)} $ and $ l_{AdS} $ is the length scale of the AdS space. Here, we place $ d = 4 $, $ l_{p} = 0.1 $, and $ a = 0.01 $ in both cases. Now, we put the radius of the cosmological horizon in Eq. (22) and Eq. (23),
	
	\begin{equation}\label{eq:area}	
		r_{CH} = \frac{1}{2} \Big[\sqrt{\eta}+\sqrt{\frac{\eta-12 M}{\eta\Lambda}}\Big],  
	\end{equation}

	\noindent and plot $ T-M $, $ T-\alpha $, $T-\Lambda$, and $ T-r $ for different values of $ M $, $ \alpha $, and $ \Lambda $ (Fig. \eqref{figure_2}).
	
	Regarding the $ T-\alpha $ plot, one can see that the Hawking temperature has been severely suppressed and decreases as the $\alpha$ coupling constant increases. Considering the $ T-M $ diagram, the temperature decreases with decreasing mass; however, the Hawking temperature rises again as the mass decreases. The behavior of the $ T-M $ can be compared with the $ T-r $ plot of Ref. \cite{37}. As expected, EUP and GEUP change Hawking temperature behavior at larger masses, in which case they predict higher Hawking temperatures. The T-M diagram illustrates that temperature behaves like temperature behavior in a flat space in small masses. $ T-\Lambda $ shows $ T_{EGB-EUP} $ and $ T_{EGB-GEUP} $ rise with the increasing cosmological constant. In this case, $ T_{EGB-GEUP} $ indicates higher Hawking temperatures at larger cosmological constants than $ T_{EGB-EUP} $.
	
	Then, we place the second root, which represents the radius of the black hole horizon, in Eq. (22) and Eq. (23)
	
	\begin{equation}\label{eq:area}	
		r_{BH} = \frac{1}{2} \Big[\sqrt{\eta}-\sqrt{\frac{\eta-12 M}{\eta\Lambda}}\Big].
	\end{equation}
	
	$ T-M $ plot shows that the Hawking temperature has decreased significantly with decreasing mass.
	The effect of $\alpha$, as the cosmological horizon, is dominant here, and increasing the coupling constant will reduce the Hawking temperature. 
	
	\begin{figure}[H]
		\centering
		\includegraphics[height=7 cm,width=7.5 cm]  {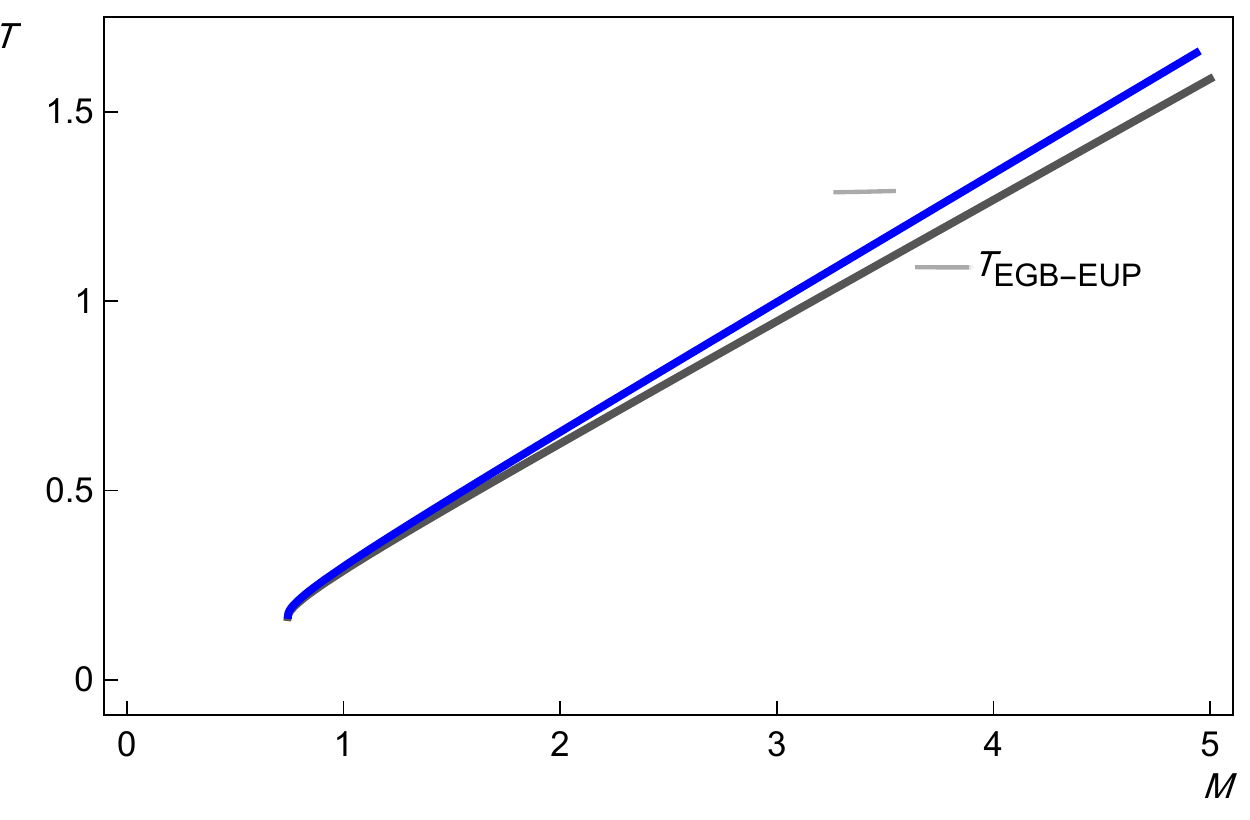}
		\hspace*{0.1cm}
		\includegraphics[height=7 cm,width=7.5 cm] {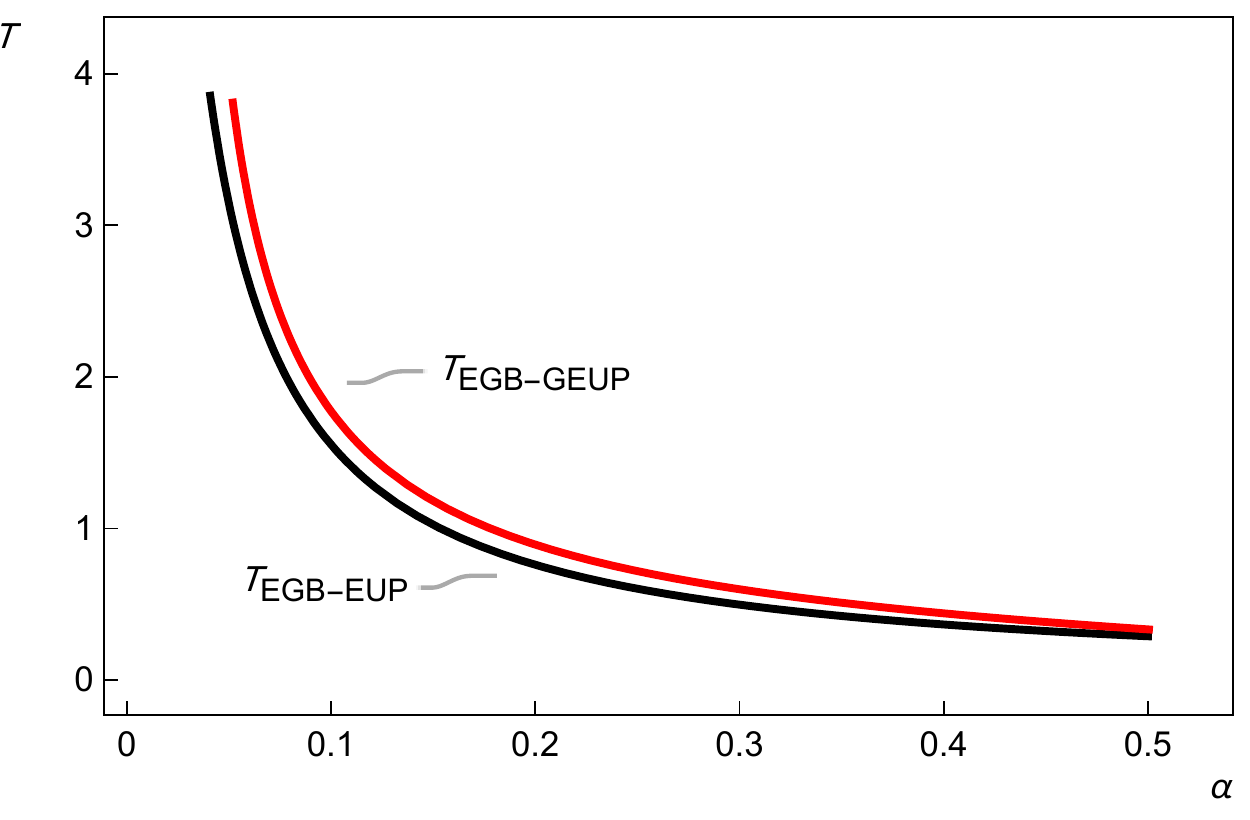}
	
		\includegraphics[height=7 cm,width=7.5 cm] {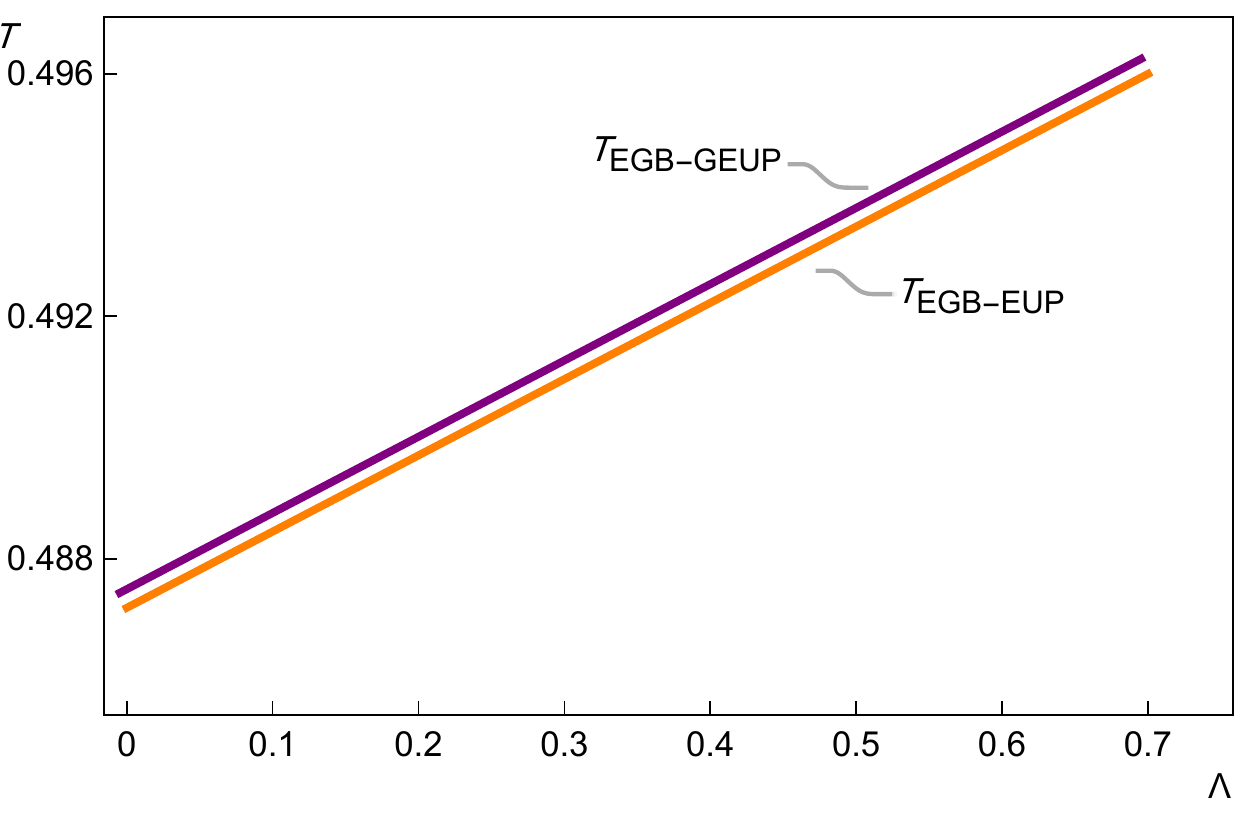}
		\hspace*{0.1cm}
		\includegraphics[height=7 cm,width=7.5 cm] {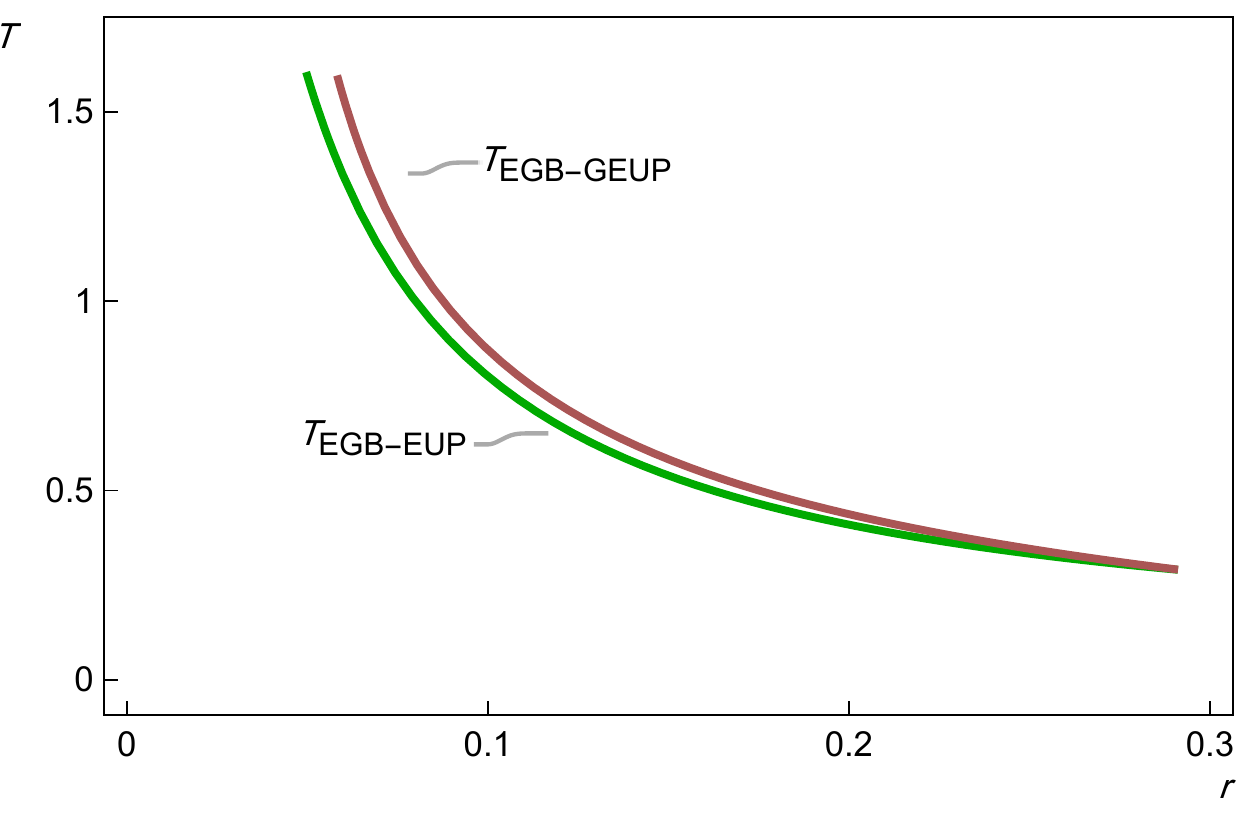}
		\caption{\scriptsize{Hawking temperature of black hole horizon versus mass, coupling constant, cosmological constant and radius for AdS 4D-EGB black hole. From the top left to right, Hawking temperature has been plotted versus different masses with $ \alpha=0.5$, and coupling constant with $M = 1$, respectively. From bottom left to right, Hawking temperature versus cosmological constant and radius has been illustrated with $ \alpha = 0.3 $. Here in all cases $ l_{AdS}=2 $, $ \Lambda=0.7 $.}}
		\label{figure_3}
	\end{figure}
	According to the $ T-\Lambda $ diagram, it can be seen that on the black hole horizon, when the cosmological constant increases, the Hawking temperature remarkably rises. In Ref. \cite{24}, the behavior of the $ T-\Lambda $ for the ``de Sitter" spacetime has been plotted. This behavior of $ T_{EGB-EUP} $ and $ T_{EGB-GEUP} $ versus $ \Lambda $ in the ``Anti-de Sitter" spacetime is the opposite of the  $ T_{EUP} $ behavior predicted in Ref. \cite{24} for ``de Sitter" spacetime.
	Furthermore, we plotted $ T-r $ for both the black hole horizon and the cosmological horizon in Fig. \eqref{figure_2} and Fig. \eqref{figure_3}. Here, $ T_{EGB-EUP} $ and $ T_{EGB-GEUP} $ with increasing radius show a decreasing behavior which is in accordance with the $ T-r $ diagram shown in Ref. \cite{37}.\\
	
	\section{Hawking temperature through EUP and GEUP in an asymptotically de Sitter 4D-EGB black hole}
	In this section, we also want to study Hawking temperature through EUP and GEUP in de Sitter spacetime. For this purpose, it is enough to put $ l^{2}=-l^{2} $ in Eq. (22) and Eq. (23)
	
	\begin{equation}\label{eq:area}	
		\Delta x_{i} \Delta p_{j} \geq \hbar \delta _{ij} \left[ 1 - \beta ^{2} \frac{(\Delta x_{i})^{2}}{l^{2}} \right],
	\end{equation}
	
	\begin{equation}\label{eq:area}	
		\Delta x_{i} \Delta p_{j} \geq \hbar \delta_{ij} \left[ 1 + a^{2} l^{2}_{P} \frac{(\Delta p_{j})^{2}}{\hbar^{2}} - \beta^{2} \frac{(\Delta x_{i})^{2}}{l^{2}} \right], 
	\end{equation}

	In de Sitter spacetime, the EUP admits an absolute maximum to $ \Delta x_{i}^{2} $, so that $  \Delta p_{i}^{2} $ is always positive 
	
	\begin{equation}\label{eq:area}	
		\Delta x_{i} \le \frac{l}{\beta},
	\end{equation}

	\noindent this upper boundary represents the Nariai boundary, where the black hole horizon and the cosmological horizon meet
	
	\begin{equation}\label{eq:area}	
		 r_{i} \le \sqrt{\frac{d-3}{d-1}} l_{ds}.
	\end{equation}
	
	At long distances in dS space, the Nariai solution is the limit of the largest black hole, which consists of two horizons: the Schwarzschild black hole horizon and a cosmological dS horizon. At the Nariai bound, the black hole becomes larger and larger until the event horizon of the black hole has an area equal to the cosmological de Sitter horizon. At this point, spacetime evolves regularly, and the singularity of the black hole tends to infinity. A spacetime symmetry in this space will connect the black hole and cosmological horizons \cite{59, 60}.
	Therefore, according to Eq. (28), there is an upper boundary for the values of the horizons, which if the value of the obtained radius is greater than this limit, the horizon of the black hole will not be formed.
	In fact, it implies that position uncertainty never transcends the cosmological horizon. 
	In addition, GEUP also creates only an absolute minimum on $ \Delta x_{i}^{2} $ in the dS space \cite{37}
	
	\begin{equation}\label{eq:area}	
		(\Delta x_{i})^{2} \geq \frac{4a^{2}l^{2}_{P}}{1+4a^{2}l^{2}_{P} \beta^{2} / l^{2}}.
	\end{equation}
	
	Furthermore, for $\Delta p_{i} $ not to be negative, a condition similar to Eq. (27) must exist. Similar to the previous sections, we apply the desired assumptions and obtain the Hawking temperature of a dS 4D-EGB black hole, with $ \Lambda=+ (d-1)(d-2)/2{l^{2}_{dS}} > 0 $,
	
	\begin{equation}\label{eq:area}	
		T_{EUP(dS)} = \left( \frac{d-3}{4\pi} \right) \hbar \left[ \frac{1}{r_{\pm}} - \left(\frac{d-1}{d-3}\right) \frac{r_{\pm}}{l^{2}_{dS}} \right],
	\end{equation}
	
	\begin{equation}\label{eq:area}	
		T_{GEUP(dS)} = \left( \frac{d-3}{4\pi}\right) \frac{\hbar r_{\pm}}{2a^{2}l^{2}_{P}} \left[ 1 - \sqrt{1 - \frac{4 a^{2} l^{2}_{P}}{r^{2}_{\pm}} \left[1- \left(\frac{d-1}{d-3} \right) \frac{r^{2}_{\pm}}{l^{2}_{dS}}\right]}\right].  
	\end{equation}
	
	The 4D-EGB black hole in dS has three horizons as follows
	
	\begin{equation}\label{horizonds}
		r=\pm\frac{1}{2} \Big[\sqrt{\eta}\mp\sqrt{\frac{\eta-12 M}{\eta\Lambda}}\Big],
	\end{equation}
	where
	\begin{equation}
		\eta=\frac{2}{\Lambda}+\frac{3\times 2^{1/3}(1-4\alpha \Lambda)}{\Lambda \xi^{1/3}}+\frac{\xi^{1/3}}{3\times2^{1/3} \Lambda},
	\end{equation}
	and
	\begin{equation}
		\xi=972 \Lambda M^2 -648 \alpha \Lambda-54+\sqrt{(972 \Lambda M^2 -648 \alpha \Lambda-54)^2-4(9-36 \alpha \Lambda)^3}.
	\end{equation}
	
	\noindent where $ r_{-} = r_{BH} $ is the black hole horizon, $ r_{+} = r_{CH} $ is the cosmological horizon, and the $ r_{0} $ is the inner horizon \cite{13}. As in the previous section, we illustrate the $ T-\alpha $, $ T-\Lambda $, $ T-M $, and $ T-r $ plots for the dS 4D-EGB black hole, and then we will examine how Hawking temperature changes with these quantities.
	
	First, we place the values of $ \beta $ and $ l_{dS} $ in Eq. (27) and obtain the maximum allowable value. By placing different values $ \alpha $, $ \Lambda $ and $ M $, we have obtained the radius of the cosmological horizon. In this case, the cosmological horizon radius resulting from the 4D-EGB metric is always more extended than the upper bound that the EUP places on $ r $. In other words, due to the 4D-EGB metric feature, the size of the cosmological horizon will always be larger than the size of dS space.
	
	In the case of the black hole horizon, we first calculate the black hole Hawking temperature through the HUP, regardless of the Planck length and large scale length. To accomplish this, it is enough to put different values of $ \alpha $ in $ r_{BH} $ and calculate $ T_{HUP} $.
	 
	According to Table. 1, we see that the dS radius of the black hole horizon decreases by increasing coupling constants. According to the $ T_{HUP} $ equation, the Hawking temperature will rise as the radius decreases. As a result, increasing the coupling constant raises the Hawking temperature.
	
	\begin{table}[H]
		\centering
		\label{table:1}
		\begin{tabular} {l|l|l}
			$ \alpha $ & $ r $ & $ T_{HUP} $  \\ \hline
			0.1 & 2.44 & 0.40  \\ 
			0.2 & 2.34 & 0.42  \\
			0.3 & 2.24 & 0.44  \\
			0.4 & 2.13 & 0.46  \\
			0.5 & 2.03 & 0.49  \\ 
		\end{tabular}
		\caption{Hawking temperature for the black hole horizon in the dS 4D-EGB black hole for different $ \alpha $ and $ r_{BH} $. Here we chose $ M = 1 $ and $ \Lambda = 0.1 $.}
	\end{table}
	
	\begin{figure}[H]
		\centering
		\includegraphics[height=7 cm,width=7.5 cm]  {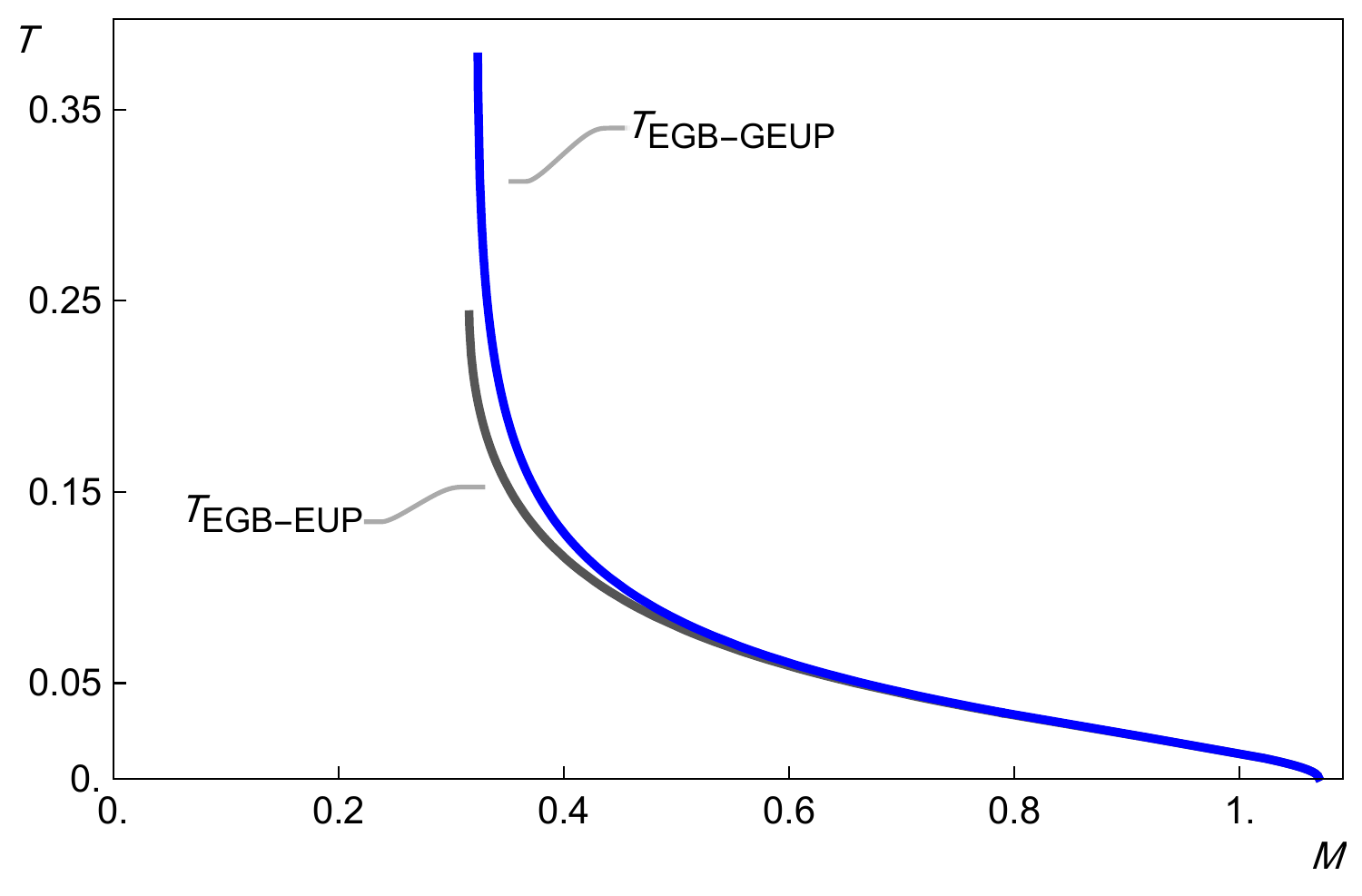}
		\hspace*{0.1cm}
		\includegraphics[height=7 cm,width=7.5 cm] {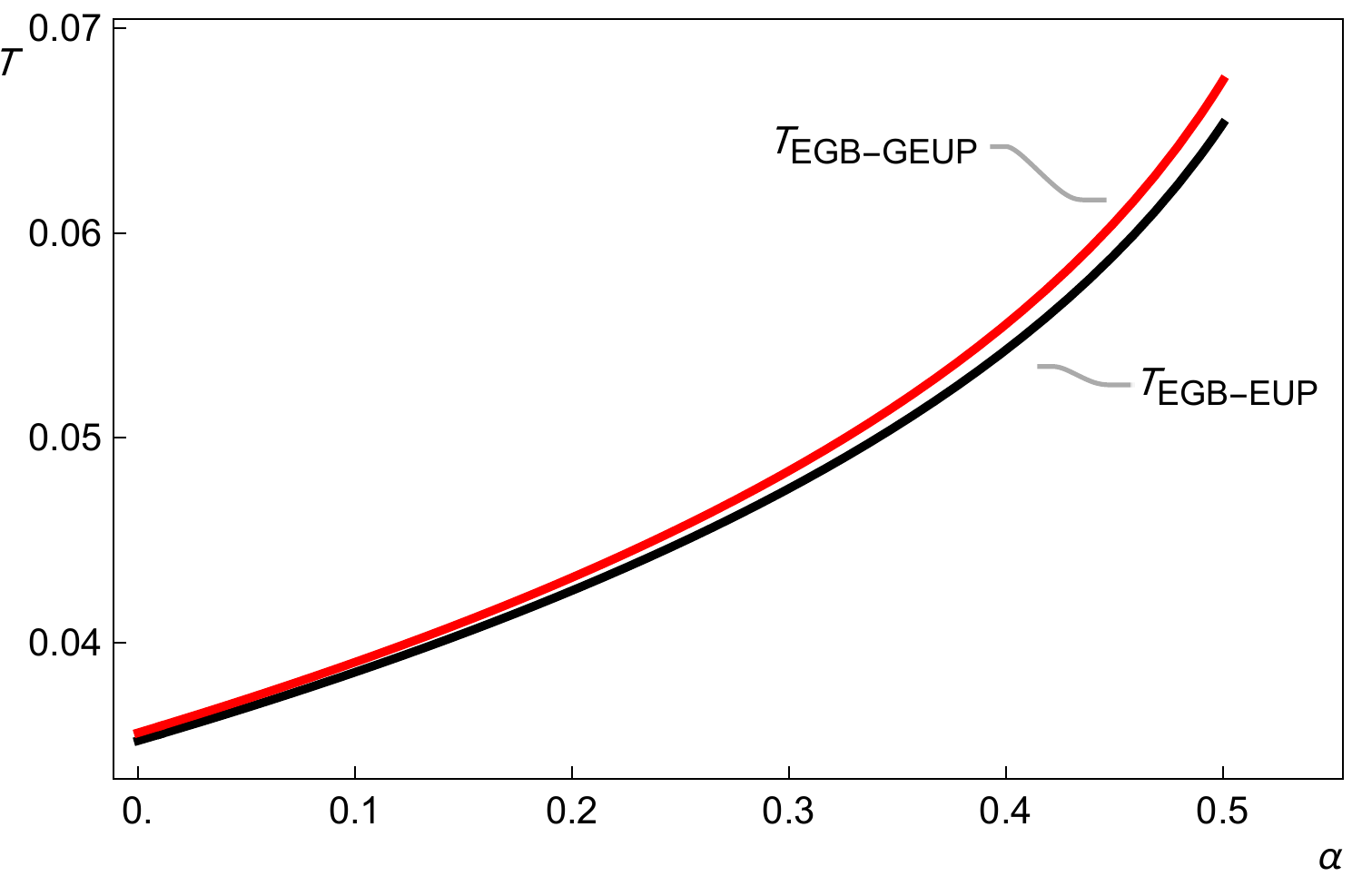}
		
		\includegraphics[height=7 cm,width=7.5 cm]  {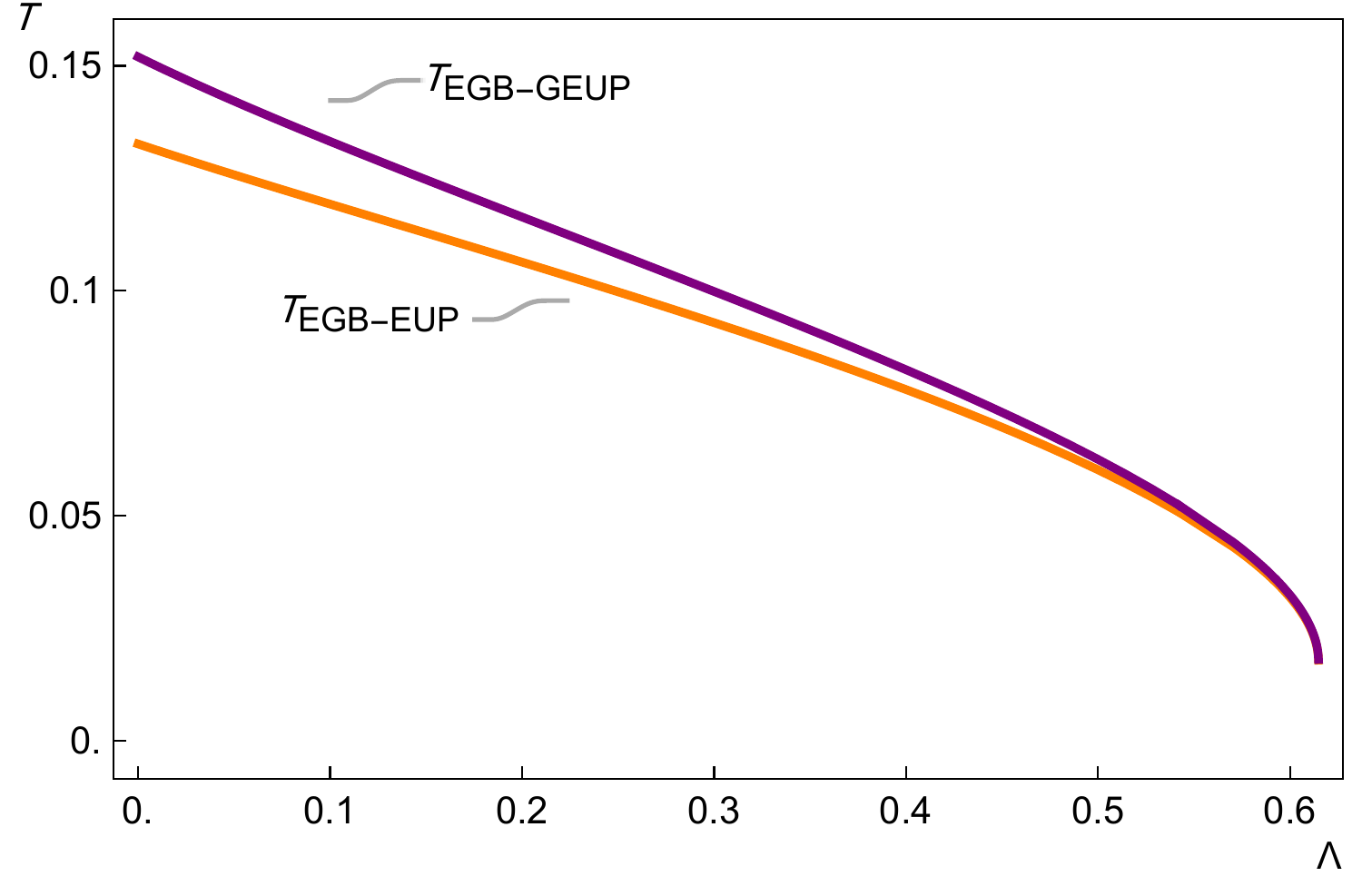}
		\hspace*{0.1cm}
		\includegraphics[height=7 cm,width=7.5 cm] {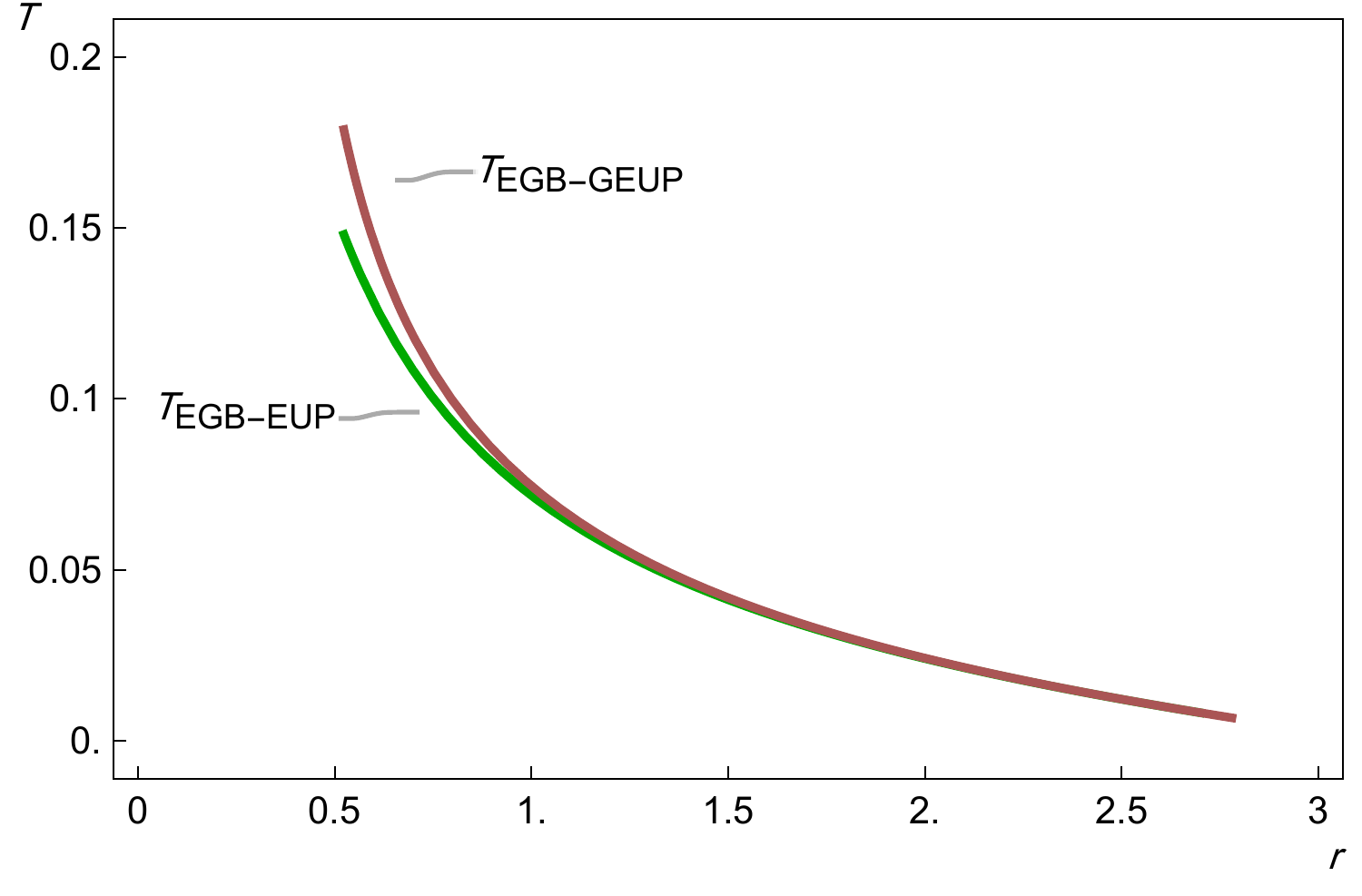}
		\caption{\scriptsize{Hawking temperature of black hole horizon versus mass, coupling constant, cosmological constant and radius for dS 4D-EGB black hole. From the top left to right, Hawking temperature has been plotted versus different masses and coupling constant with $ \alpha=0.1$ and $ M = 1$, respectively. From bottom left to right, Hawking temperature versus cosmological constant and radius has been illustrated with $ \alpha = 0.1 $, receptively. Here in all cases $ \hbar = l_{p} = 1$, $ l_{dS}=5.5 $, $ \Lambda=0.1 $, and $ a = 0.2 $.}}
		\label{figure_4}
	\end{figure}
	 
	 Regarding the Fig. \eqref{figure_4}, in contrast to the Anti-de Sitter spacetime, the $T- \alpha $ diagram for the black hole horizon shows that as the $ \alpha $ increases positively, $ T_{EGB-EUP} $ and $ T_{EGB-GEUP} $ also rise. The behavior of $ T_{EGB-EUP} $ and $ T_{EGB-GEUP} $ are similar to the changes of $ T_{HUP} $ versus $ \alpha $. 
	 However, when we plot the $ T-r $ diagram for the radius of the black hole horizon, the temperature shows an increasing behavior as the radius decreases, and the temperature is limited at one point and stops increasing. This behavior of the $ T-r $ diagram is utterly consistent with the results obtained in Ref. \cite{37}. 
	 
	 The $ T-M $ plot predicts higher Hawking temperature than flat and AdS spaces. $ T_{EGB-GEUP} $ has reached the final temperature of evaporation sooner. Indeed, having a minimum length in GEUP makes the black hole evaporation faster than when there is no minimum length and ends at a certain point \cite{37}. If, as in Ref. \cite{13}, we considered only the EGB effect, and we would expect the black hole to continue evaporation until the temperature drops to zero. Nevertheless, when we incorporate EUP and GEUP into our equations, the expected behavior in  Ref. \cite{13} will change. In this case, when the Hawking temperature reaches a limited value, evaporation no longer continues.
	 The black hole is expected to continue evaporating at EGB-EUP, but here $ T_{EGB-EUP} $ has also stopped at a certain point. This is due to the presence of EGB. The effect of EGB is more noticeable on the black hole horizon and has a more significant effect on the final temperature of the dS 4D-EGB black hole. EGB modifies the structure of the black hole and specifies a minimum mass for it. If the mass is less than this minimum mass, the black hole will no longer be considered in the mass range we are studying, i.e., this value will be less than the mass that three dS black hole horizons should have; this is why $ T_{EGB-EUP}$ has been discarded in this case. However, $ T_{EGB-EUP} $ and $ T_{EGB-GEUP} $ behave similarly in substantial masses.
	 Moreover, the $ T-\Lambda $ diagram behaves consistently with Ref. \cite{24}. Increasing the cosmological constant at different $ \alpha $ will cause $ T_{EGB-EUP} $ and $ T_{EGB-GEUP} $ to drop.\\

	\section{Conclusion}
	
	In previous studies, the EGB solution was assumed to be trivial in four dimensions, but recently the existence of 4-dimensional Einstein-Gauss-Bonnet black holes has been investigated by Glavan and Lin. EGB black holes have different properties in 4-dimension. Hawking radiation and temperature can also be expected to have unique properties compared to higher dimensions in this case. In Ref. \cite{37}, it is shown that, in general, the minimum length generated by GUP can increase the Hawking temperature, regardless of whether the space is asymptotically flat or (Anti)- de Sitter. In fact, the presence of the expression $ "a^{2} l_{P}/r_{\pm}^{3}" $ in the temperature equation can be considered a universal trend that causes the black hole to evaporate faster. We investigated the temperature behavior of the 4D-EGB black hole by using GUP, EUP, and GEUP. 
	By plotting the temperature in terms of coupling constants, cosmological constants, mass, and radius, we estimated the temperature behavior in terms of their variations. The 4D-EGB black hole has two horizons in flat space: the black hole and the white hole horizon. On the black hole horizon, temperature exhibits an increasing behavior relative to the increment in the coupling constant.
	In addition, the T-M plot illustrates that when the GUP enters the EGB, it prevents the evaporation of the black hole where the temperature has not yet reached zero.
	
	In the AdS space, the coupling constant on both the black hole and the cosmological horizons significantly affects the Hawking temperature and decreases acutely with increasing positive $ \alpha $. Hawking temperature on the black hole horizon shows increasing behavior with an increasing cosmological constant, contrary to temperature behavior in the dS space.
	On the cosmological horizon, one can see that an increase in  $ \Lambda $ increases the Hawking temperature. When plotting the Hawking temperature obtained from the EUP and GEUP for the black hole horizon versus mass, we see that the temperature drops sharply to zero when the mass decreases. On the cosmological horizon, temperature shows the same behavior, except that when the mass becomes very small, the temperature rises again. $ T_{EGB-GEUP} $ predicts a higher temperature than the $ T_{EGB-GEUP} $, and at a particular point, due to its minimum length, it stops evaporation and prevents Hawking temperature divergence.
	
	There are expected to be three horizons in the 4D-EGB black hole dS space: the black hole horizon, the cosmological horizon, and the inner horizon. In the dS space, EUP and GEUP predict an absolute maximum for position uncertainty. Due to the Nariai limit, the size of the horizons will be limited and should not exceed a specific limit. However, the size of our expected 4D-EGB black hole cosmological horizon will always be more significant than this bound, so the cosmological horizon will not be formed under this condition. 

	Ultimately, when we plot $ T-r $ for the flat, (A)dS and dS spaces, one can see that Hawking's temperature behavior versus radius is entirely in line with the predictions. GUP and GEUP affect temperature behavior and prevent it from diverging in a small radius.
	In fact, the smaller the radius, the greater the effect of GUP and GEUP on the temperature of the black hole, causing the temperature to rise sharply and evaporation to accelerate.
	Finally, the minimum length stops the black hole from evaporating at a certain point, preventing Hawking temperature divergence.\\

	
	
	
	
	

\end{document}